\documentclass{appolb}
\usepackage{epsfig}

\begin{document}
\title{The environmental effects in the origin of angular momenta of galaxies}

\author{W{\l}odzimierz God{\l}owski
\address{Institute of Physics, Opole University, Opole, Poland}
\and
Elena Panko
\address{Kalinenkov Astronomical Observatory, Nikolaev State
University, Nikolaev, Ukraine}
\and
Piotr Flin
\address{Institute of Physics, Jan Kochanowski University, Kielce, Poland}
}

\maketitle
\begin{abstract}
We study the galaxy alignment in the sample of very rich Abell
clusters located in and outside superclusters. The statistically
significant difference among investigated samples exists. We found
that in contrast to whole sample of cluster, where alignment
increase with the cluster richness, the cluster belonging to the
superclusters does not show this effect. Moreover, the alignment
decreased with the supercluster richness. One should note however
that orientations of galaxies in analyzed clusters are not random,
both in the case when we analyzed whole sample of the clusters and
only clusters belonging to the superclusters. The observed trend,
dependence of galaxy alignment on both cluster location and
supercluster richness clearly supports the influence of
environmental effects to the origin of galaxy angular momenta.
\end{abstract}

\section{Introduction}

The problem of the origin of large scale structures is till now one
of the most enigmatic ones. It is commonly accepted that presently
observed structures originated from almost isotropic distribution in
the early Universe. The departure from isotropy, as estimated from
CMBR is of the order of $\delta\rho/\rho\simeq 10^{-5}$. About a
half century ago the main problems were connected with type of the
perturbations, its amplitude and scale (mass or length). Different
theories of galaxy origin called scenarios predicted various masses
of newly originate structures and various manners in which galaxy
gained angular momentum (Peebles 1969, Zeldovich 1970, Sunyaev \&
Zeldovich 1972, Doroshkevich 1973, Shandarin 1974, Wesson 1982, Silk
\& Efstathiou 1983, Bower et al. 2005). In some scenarios, the
distribution of galaxy angular momenta in structure were random
ones, in others perpendicular or parallel to the structure main
plane (Peebles 1969, Doroshkevich 1973, Shandarin 1974, Silk 1983,
Catelan \& Theuns 1996, Li 1998, Lee \& Pen 2002, Navarro et al.
2004, Trujillio et al. 2006). In such a manner, the existence or not
of the galaxy orientation can be used for testing scenarios of
galaxy origin. Because the real location of rotation axes is known
for very small number of objects, usually the study of orientation
of galaxy planes is performed. In contemporary picture of large
scale distribution known as "Cosmic Web" we practically have four
components. These are long filaments, walls, voids and the rich,
dense regions called galaxy clusters.

In our previous paper (God{\l}owski \& Flin 2010) we studied the
orientation of galaxy groups in the Local Supercluster. We found the
strong alignment of the major axis of the groups with both direction
toward the supercluster centre (Virgo cluster) as well as with line
joing two brightest galaxies in the group. The interpretation of
these observational facts was the following. The brightest (belived
to be the most massive) galaxies of the group originate first. Due
to gravitational forces other galaxies are attracted to these ones
and a filament is forming. The other possibility is that at the pre-
existing filament galaxies are forming. In this direction is going
the result of Jones et al. (2010) which interpretet theirs founding
that the spins of spiral galaxies located within cosmic web
filaments tend to be aligned along the larger axis of the filament,
as "fossil" evidence indicating that the action of large scale tidal
torques effected the alignments of galaxies located in cosmic
filaments.

In the paper of God{\l}owski et al. (2010), when analizing the
sample of 247 rich Abell Clusters, we found that the alignment of
member galaxies in rich structures, having more than 100 members, is
a function of the group mass. Richer group exhibits stronger galaxy
alignment. The change of alignment with the surrounding neigbourhood
was observed also in alignment study in void vicinity (Varela et al.
2011), being continuation of earlier study of galaxy orientation in
regions surrounding bubble-like voids (Trujillo et al. 2006). They
found that disk galaxies around large voids (greater than $15
Mpc/h$) present a significant tendency to have their angular momenta
aligned with the void's radial direction. The strength of this
alignment is dependent on the void's radius and for voids with $\le
15 Mpc/h$ the distribution of the orientation of the galaxies is
compatible with a random distribution. Varela et al. (2011) found
that this trend observed in the alignment of galaxies is similar to
that observed in numerical simulations of the distribution of dark
matter i.e. in distribution of the minor axis of dark matter halos
around cosmic voids, which suggests a possible link in the evolution
of both components.

In view of these facts, it is interesting to look if the the cluster
belonging to the larger structures exhibit the same type of
alignment as whole sample of the clusters. For this reason, we
analize the alignment of the cluster members of galaxies for the
clusters belonging to the superclusters. This problem was not
investigated till now, although aligment of galaxies in the
superclusters was investigated many times. Presence of non-random
galaxy spin orientation has been ascertained both in the Local
Supercluster (for example MacGillivray et al. 1982, Flin and
God{\l}owski 1986, 1989, 1990, God{\l}owski 1993, 1994, Kashikawa
and Okamura 1992, Aryal and Saurer 2005a, Hu et al. 2006,
 Aryal, Neupane and Saurer 2008, Aryal, Paudel
and Saurer 2008) and in other superclusters, as the Hercules
Supercluster, Coma/A 1367 and the Perseus Supercluster (Gregoryet
al. 1981, Djorgovski 1983, Flin 1988, Garrido et al. 1993, Wu et al.
1997, Cabanela and Aldering 1998, Flin 2001).

\begin{figure}[!htb]
\centerline{\psfig{file=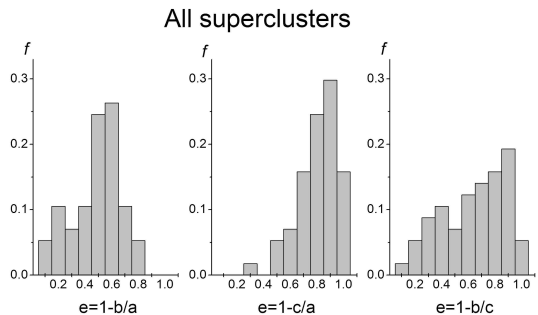,width=0.8\textwidth}}
\vspace*{8pt} \caption{Ellipticities of the analyzed superclusters
\label{fig:f1}}
\end{figure}

\section{Observational data }

The PF Catalogue of galaxy structures (Panko \& Flin 2006) was used
as observational basis of the present study. This Catalogue was
constructed by finding structures in the Muenster Red Sky Survey
(MRSS) (Ungruheet al. 2003). We used the Voronoi tessellation
technique for structure finding. MRSS is a large-sky galaxy
catalogue covering an area of about 5000 square degrees on the
southern hemisphere. It is the result of scanning of 217 ESO plates,
giving positions, red magnitudes, radii, ellipticities and position
angles of about 5.5 million galaxies and it is complete down to
$r_F=18.3^m$. As a result we have 6188 galaxy structures called
clusters. Using standard covariance elipse method we determined
structure ellipticity and position angles. We chose the sample of
247 very rich clusters, having at least 100 members each, and being
identified with one of the ACO clusters (Abell et al. 1989)- see
God{\l}owski et al. (2010) for details. The PF catalogue served also
as a basis for supercluster detection. We found 54 superclusters
having at least 4 clusters each. As expected, superclusters are
rather flat structures (see Fig. \ref{fig:f1}). We found that $43$
from total member of $247$ clusters belong to the superclusters and
they were chosen for detailed analyzis.

\begin{table}[h]
\begin{center}
\scriptsize \caption{The frequency of anisotropy of very rich
clusters located in superclusters}
\label{tab:t1}
\begin{tabular}{cccc}
\hline
Richness     &    The angle P& The angle $\delta_D$&  The angle $\eta$\\
\hline
   N=4           &      0.84   &          0.74  &      0.84\\
   N=5-7         &      0.31   &          0.90  &      0.79\\
   N=8-10        &      0.43   &          0.57  &      0.43\\
\hline
\end{tabular}
\end{center}
\end{table}

\section{Results and analysis}

We studied the alignment of galaxies in very rich clusters belonging
to superclusters. The study of alignment as usually was done by
analyzing the angles connected with the orientation of galaxy plane.
These are: the position angle of the great axis of the galaxy image
and the angles describing the orientation of the normal to the
galaxy planes: polar - $\delta_D$ and azimuthal $\eta$. At first we
binned our samples into three bins according to the supercluster
richness, These were: superclusters containing only 4 structures,
subsample containing 5,6 and 7 structures and finally subsample of
superclusters each of them containing 8 and 10 clusters. One should
note hovewer that three clusters 0347-5571, 2217-5177 and 2234-5249
have double possible identification with supercluster and these
clusters are counted for two bins.

\begin{table}[h]
\begin{center}
\scriptsize \caption{ The random statistics} \label{tab:t2}
\begin{tabular}{ccccc}
\hline
Test&$\bar{x}$&$\sigma(x)$&$\sigma(\bar{x})$&$\sigma(\sigma(x))$\\
\hline
  $\chi^2$                       &  34.9592 &  1.2843  &  0.0406 &  0.0287\\
  $\Delta_{1}/\sigma(\Delta_{1})$&   1.2567 &  0.0983  &  0.0031 &  0.0022\\
  $\Delta/\sigma(\Delta)$        &   1.8846 &  0.1027  &  0.0032 &  0.0023\\
\hline
\end{tabular}
\end{center}
\end{table}

\begin{table}[h]
\begin{center}
\scriptsize
\caption{The statistics of the observed distributions
for real clusters} \label{tab:t3}
\begin{tabular}{c|cc|cc|cc}
\hline
\multicolumn{1}{c}{}& \multicolumn{2}{c}{$P$}&
\multicolumn{2}{c}{$\delta_D$}&
\multicolumn{2}{c}{$\eta$}\\
\hline
Test&$\bar{x}$&$\sigma(x)$&$\bar{x}$&$\sigma(x)$&$\bar{x}$&$\sigma(x)$\\
\hline
  $\chi^2$                       &  38.772 &  1.574&57.079&6.190&84.656&7.391 \\
  $\Delta_{1}/\sigma(\Delta_{1})$&   1.797 &  0.148& 3.594&0.385& 5.324&0.586 \\
  $\Delta/\sigma(\Delta)$        &   2.339 &  0.148& 4.906&0.407& 6.475&0.522 \\
\hline
\end{tabular}
\end{center}
\end{table}

\begin{figure}[b]
\begin{center}$
\begin{array}{lll}
\includegraphics[angle=270,scale=0.14]{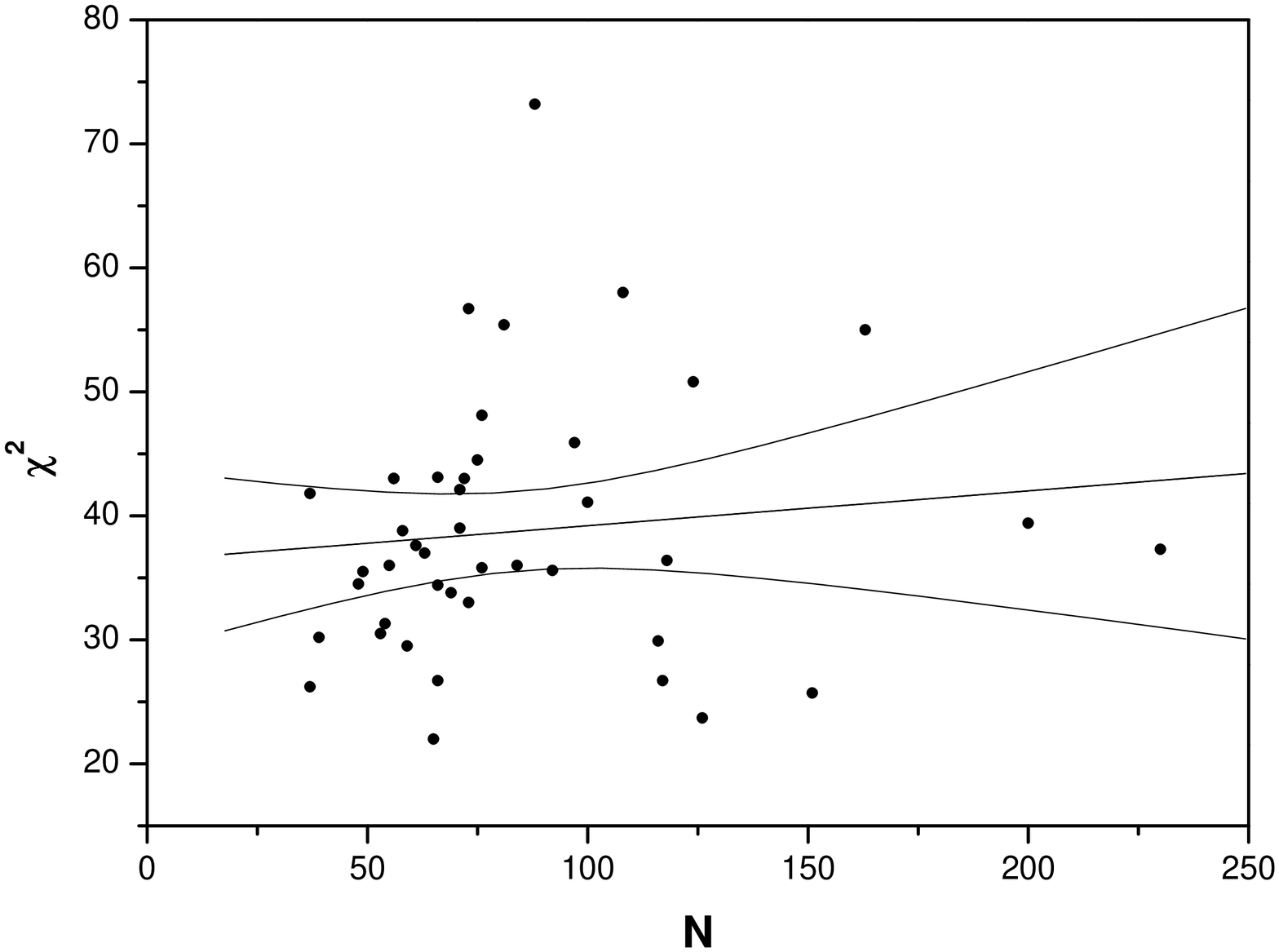} & \includegraphics[angle=270,scale=0.14]{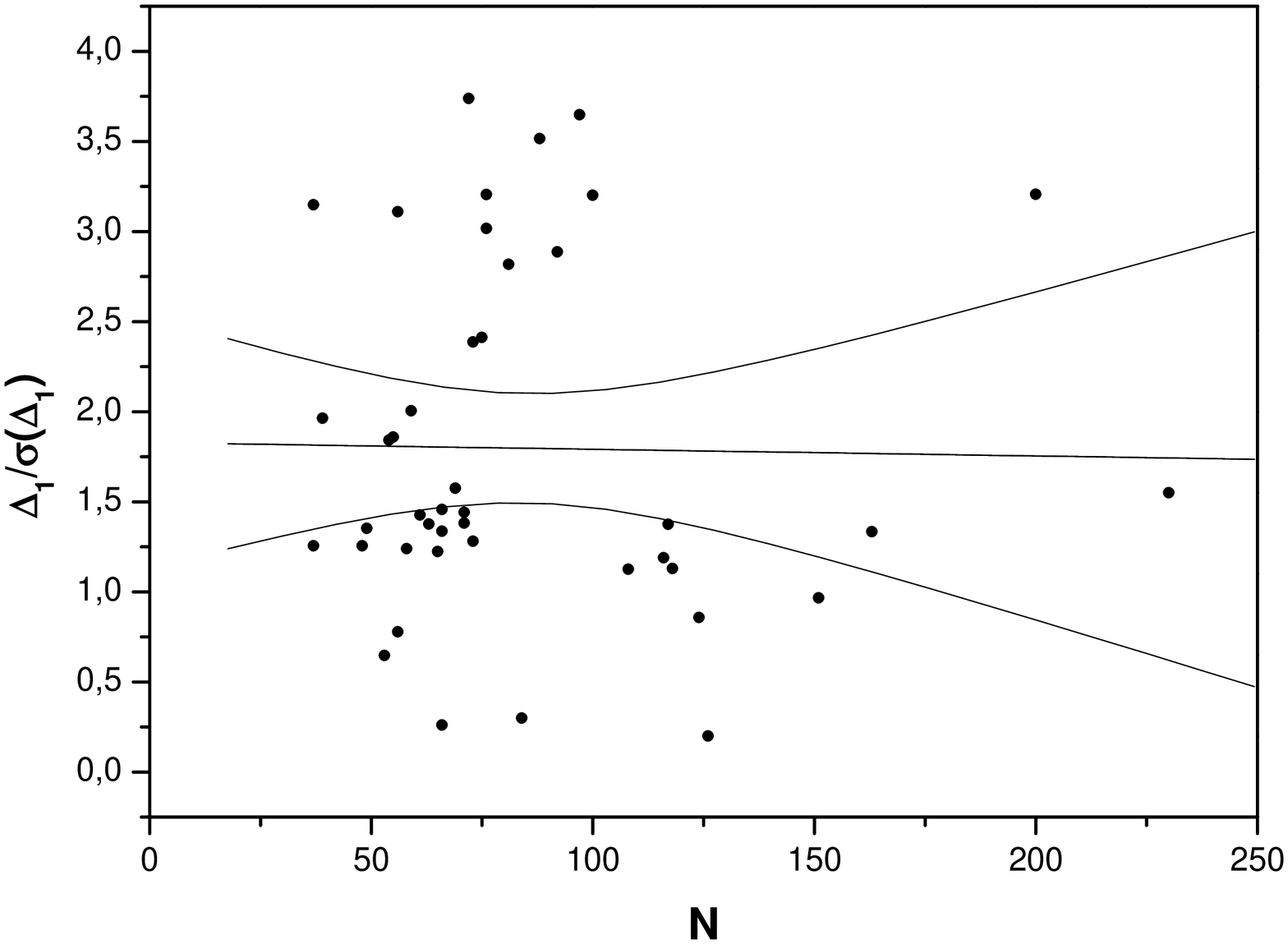} & \includegraphics[angle=270,scale=0.14]{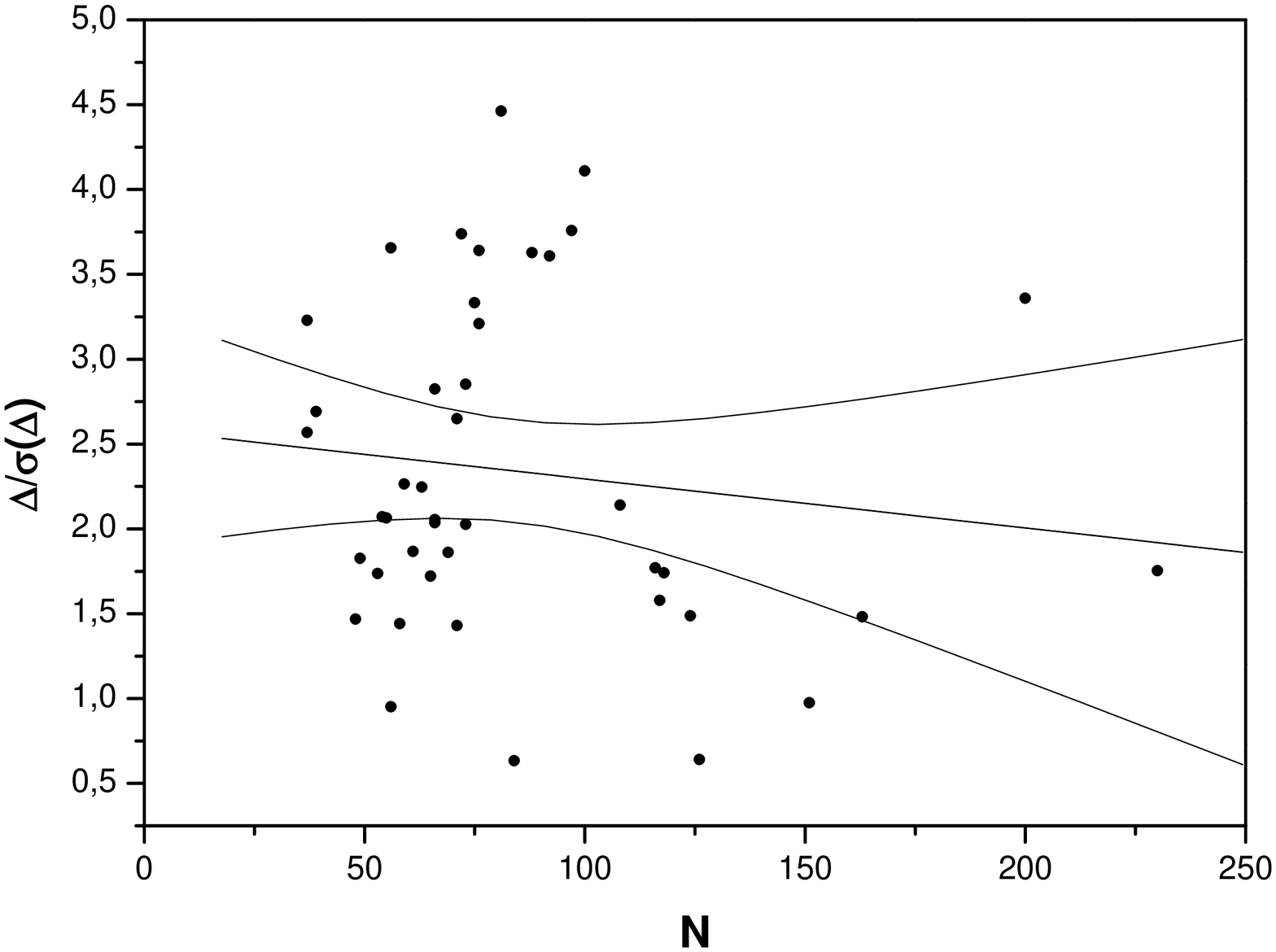}\\
\includegraphics[angle=270,scale=0.14]{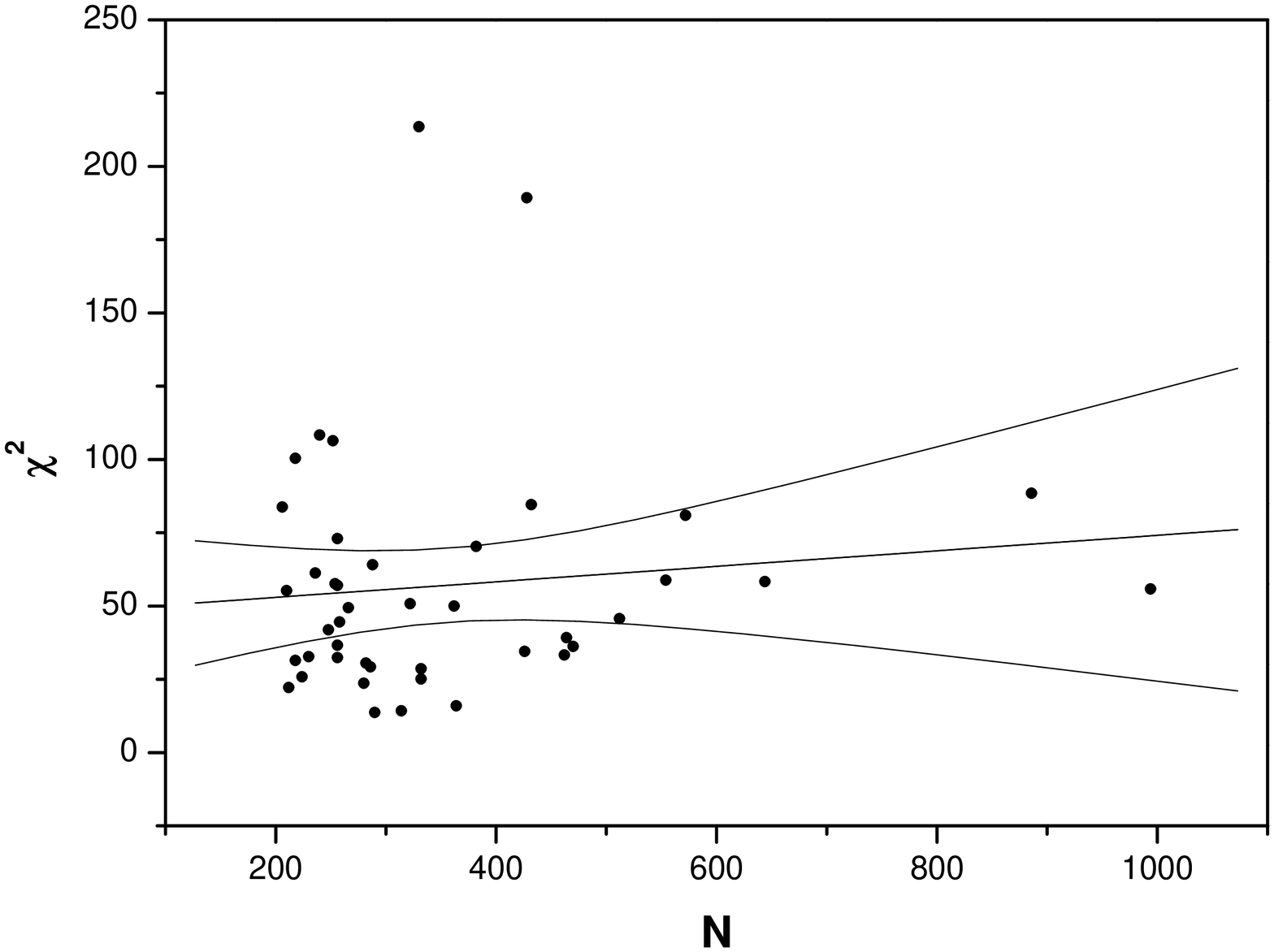} & \includegraphics[angle=270,scale=0.14]{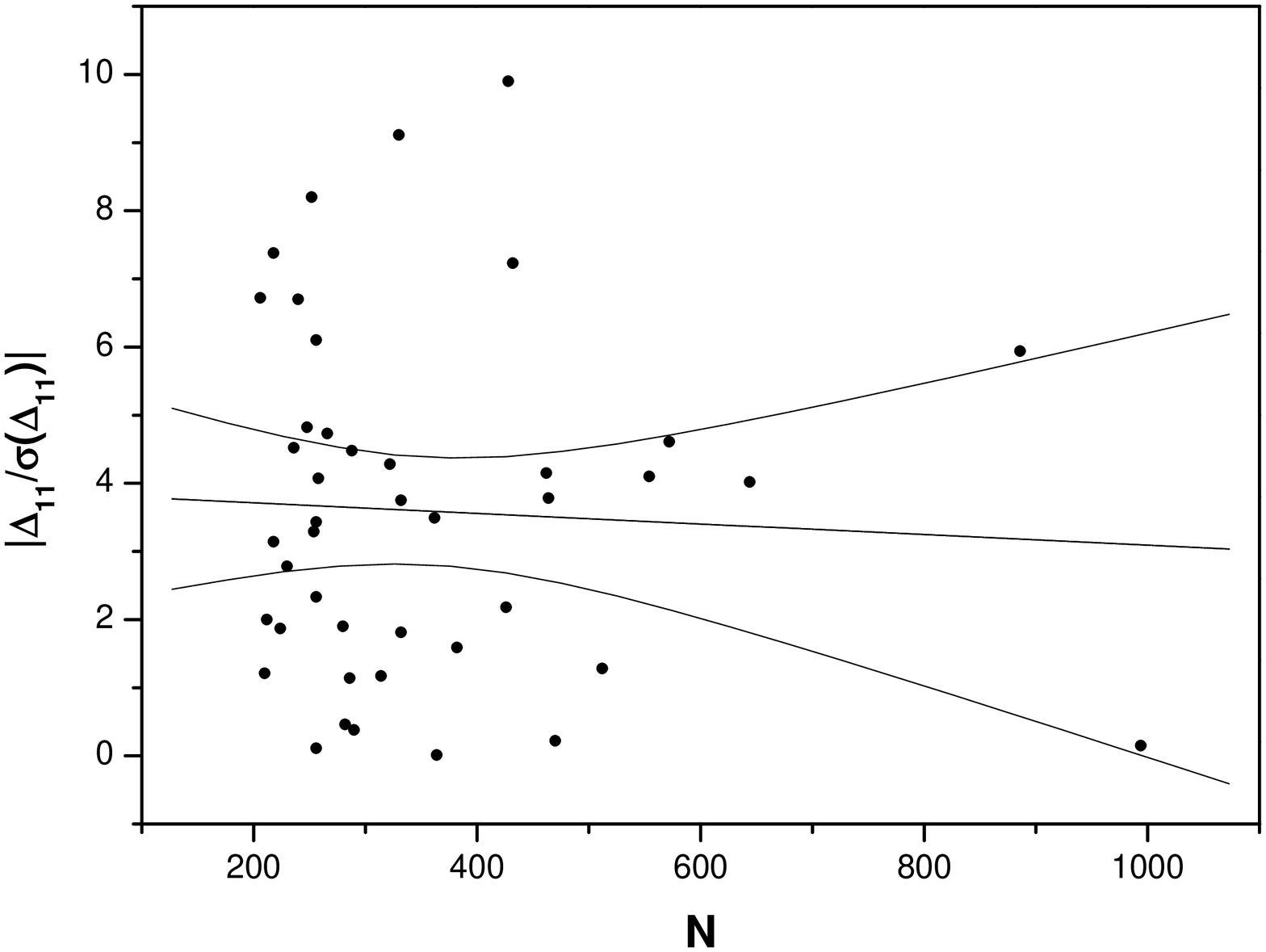} & \includegraphics[angle=270,scale=0.14]{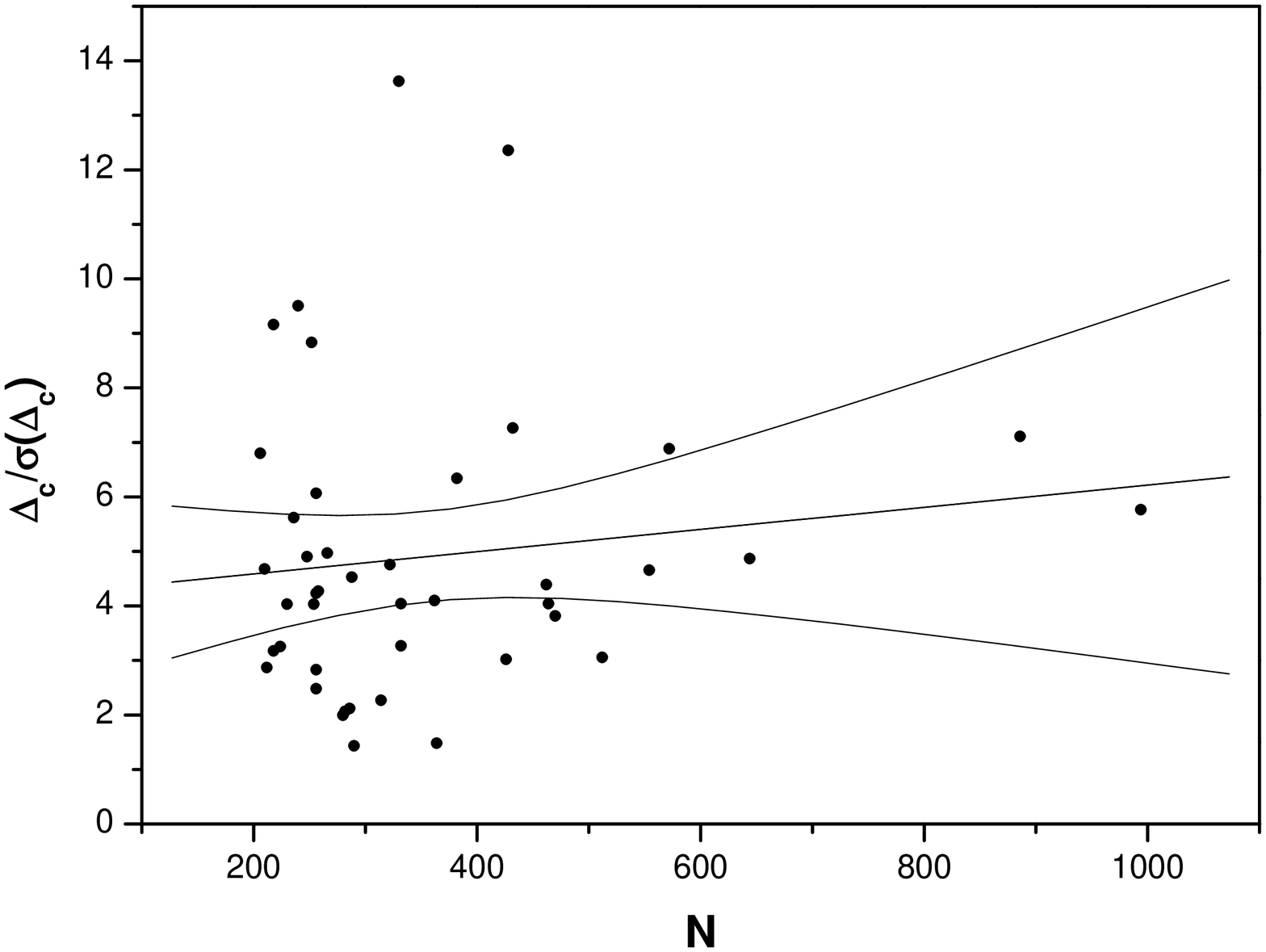}\\
\includegraphics[angle=270,scale=0.14]{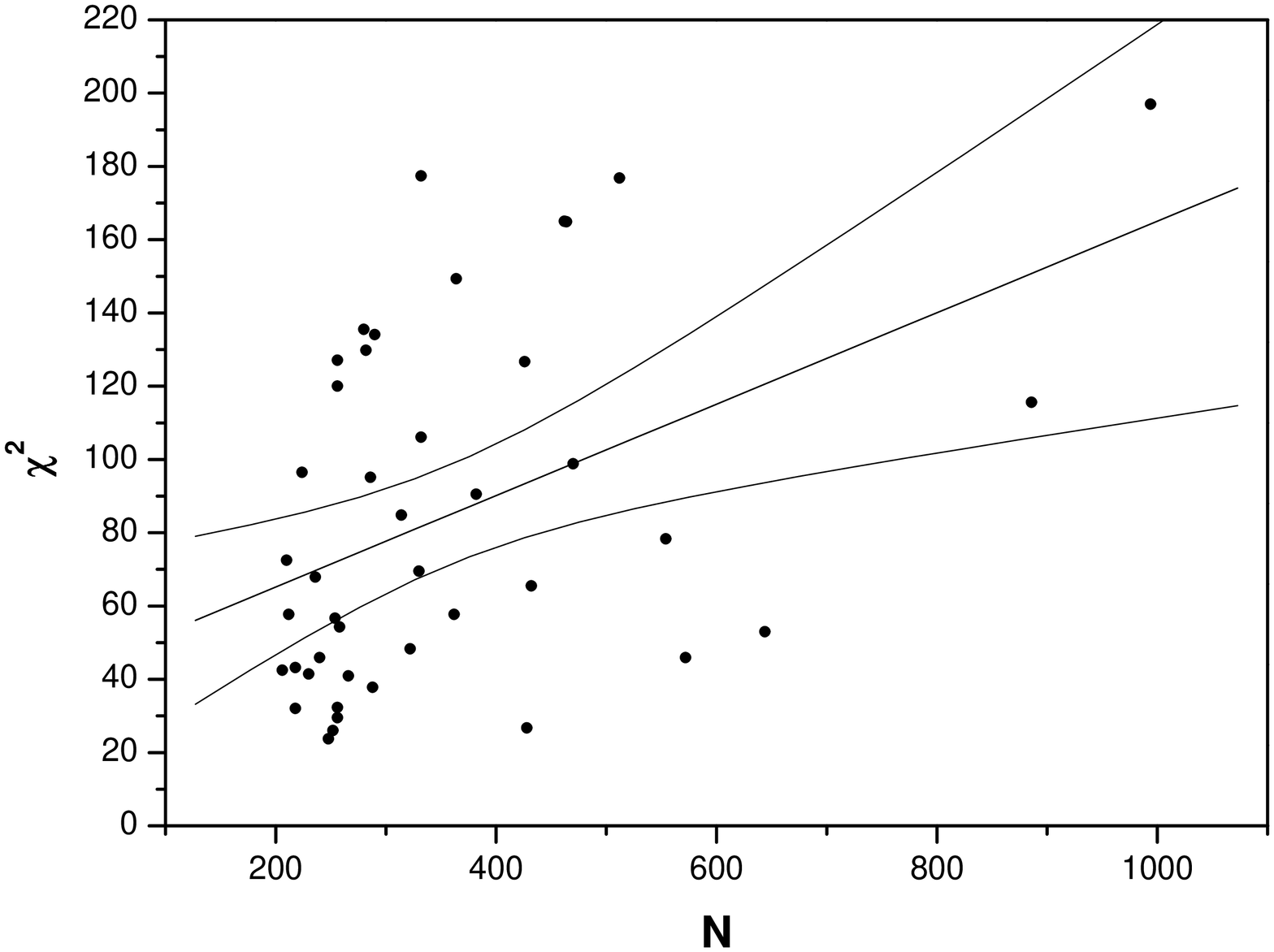} & \includegraphics[angle=270,scale=0.14]{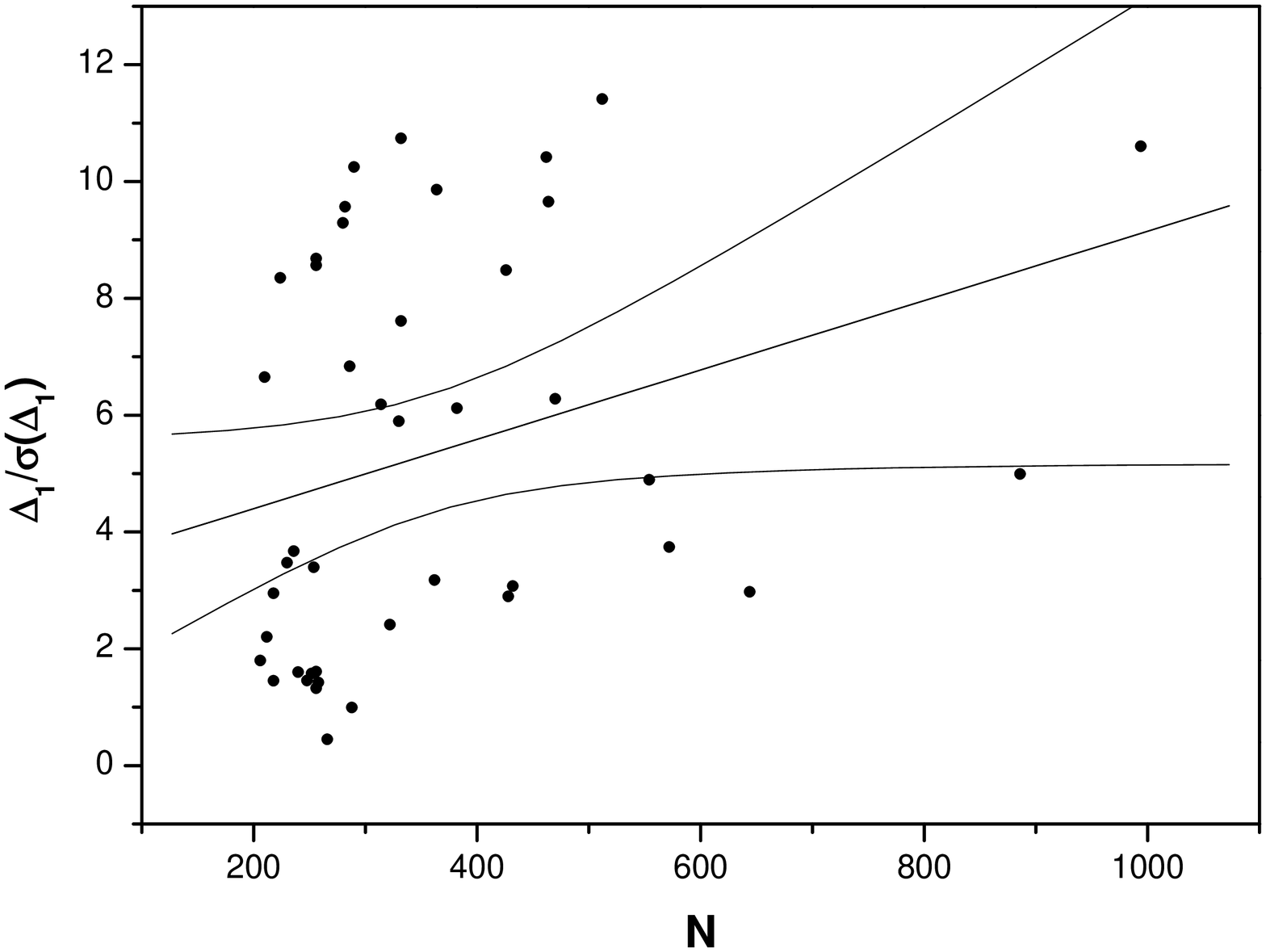} & \includegraphics[angle=270,scale=0.14]{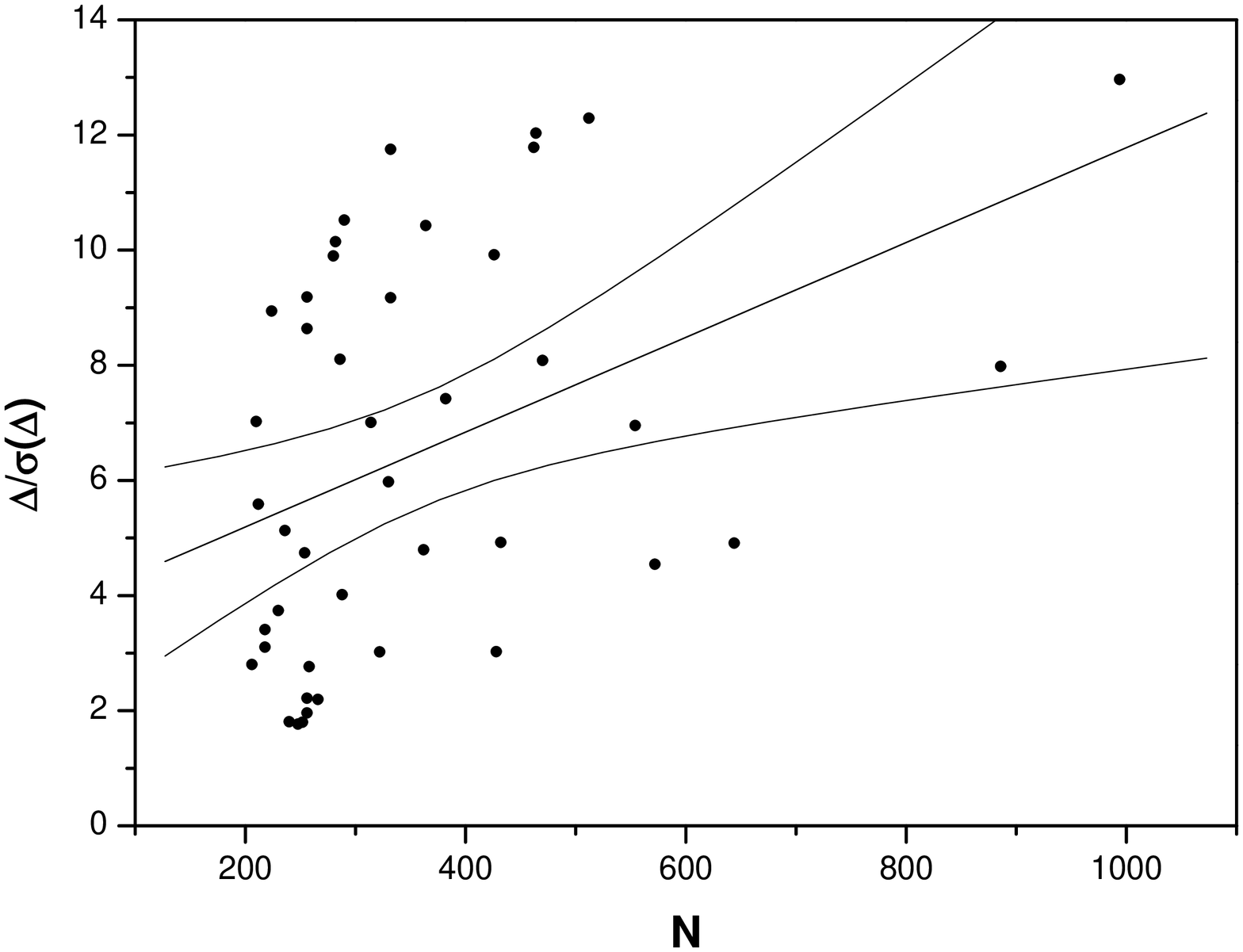}\\
\end{array}$
\end{center}
\caption{The relation  between the number of galaxies in the cluster
$N$ and the value of analyzed statistics ($\chi^2$ - left panel,
$\Delta_1/\sigma(\Delta_1)$ - middle panel, $\Delta/\sigma(\Delta)$
- right panel)  for the position angles $p$ (upper panel),$\delta_D$
angles (midlle panel) and $\eta$ angles (bottom panel) - equatorial
coordinates. The bounds error, at confidence level $95\%$, were
presented as well.
\label{fig2}}
\end{figure}

At first we study the frequency of alignment of very rich clusters
(having 100 and more members) ascribed to supercluster (Tab.1). One
can show that anisotropy decreased with supercluster richness.
Moreover, we determined the frequency alignment in the whole sample
of 247 very rich Abell clusters. In the whole sample of $247$
clusters anisotropy on the distributions of the angle P was observed
in $38\%$ clusters, while galaxy plane anisotropy was noted in
$85\%$ and $77\%$ clusters in the case of the angle $\delta_D$ and
the angle $\eta$ respectively. For the sample of galaxies belonging
to the supeclusters we observed anisotropy in $55\%$ clusters in the
case of position angles $P$ and in $78\%$ and $76\%$ of the clusters
in the case of the angle $\delta_D$ and the angle $\eta$
respectively.

The observed anisotropy is significantly greater when we analyzed
the spatial orientation of galaxy plane ($\delta_d$ and $\eta$
angles) than in the case of position angles $p$. In our opinion this
is due to incorrectly assumed shapes of galaxies. This problem was
analyzed in details by God{\l}owski,Ostrowski (1999), God{\l}owski
et al. (1998), Baier et al. (2003). The work of God{\l}owski and
Ostrowski (1999) was based on the Tully's NGC Catalog (1988). In
this catalog during calculating galaxy inclination angles, Tully
assumed that the "true" ratio of axes of galaxies is 0.2, which is a
rather poor approximation, especially for non-spiral galaxies
(God{\l}owski 2011). For our present analyzis this efect is not so
important because in the case of analisis of spatial orientation of
galaxy planes our interest is how aligment is changing with
belonging of the clusters to the supercluster and with supercluster
richness.

Now we decided to analyze the alignment in the cluster belonging to
superclusters in more details. Hawley \& Peebles (1975) analyzed the
distributions of position angles using $\chi^2$ test, Fourier tests
and autocorrelation test. Since Hawley \& Peebles (1975) paper this
method was accepted as standard method for analysis of a galactic
alignment. One should note that there are several modifications and
improvement of original Hawley \& Peebles (1975) methods (Flin \&
God{\l}owski 1986, Kindl 1987, God{\l}owski 1993, 1994, Aryal \&
Saurer 2000, God{\l}owski et al.2010) God{\l}owski (2011a) performed
a deeper improvement of the original Hawley \& Peebles (1975) method
and showed its usefulness for analysis of galactic orientations in
clusters. In the paper of God{\l}owski (2011a) the mean values of
analyzed statistics was computed. The null hypothesis $H_0$ was that
the mean value of the analyzed statistics was as expected in the
cases of a random distribution of analyzed angles. The results was
compared with theoretical predictions as well as with results
obtained from numerical simulations.

\begin{table}[h]
\begin{center}
\scriptsize \caption{The results of the linear regression analysis
 - value of analyzed statistics on function of the cluster richness for the clusters belongin
to the superclusters. \label{tab:t4}} \resizebox{\textwidth}{!}{
\begin{tabular}{c|cc|cc|cc}
\hline \multicolumn{1}{c}{}& \multicolumn{2}{c}{$\chi^2$}&
\multicolumn{2}{c}{$\Delta_1/\sigma(\Delta_1)$}&
\multicolumn{2}{c}{$\Delta/\sigma(\Delta)$}\\
\hline
angle&$a \pm\sigma(a)$&$b\pm\sigma(b)$&$a\pm\sigma(a)$&$b\pm\sigma(b)$&$a\pm\sigma(a)$&$b\pm\sigma(b)$\\
\hline
$P$     &$0.028\pm 0.039$&$36.4\pm3.7 $&$-.0004\pm0.0037$&$1.83\pm0.15$&$-.0029\pm0.0037$&$2.58\pm0.34$\\
$\delta$&$0.026\pm 0.037$&$47.7\pm14.6$&$-.0008\pm0.0023$&$3.87\pm0.91$&$0.0020\pm0.0024$&$4.18\pm0.96$\\
$\eta$  &$0.125\pm 0.040$&$40.2\pm15.7$&$0.0060\pm0.0030$&$3.21\pm1.17$&$0.0082\pm0.0028$&$3.54\pm1.13$\\
\hline
\end{tabular}}
\end{center}
\end{table}

Following God{\l}owski (2011a) method now we analyzed our sample of
$43$ clusters belonging to the superclusters in details. Because of
small number of galaxies in some clusters we performed 1000
simulations of the distributions of the position angles in 43
fictious clusters, each cluster with number of members galaxies the
same as in the real cluster. In the Tab.2 we present average values
of the analyzed statistics, theirs standard deviations, standard
deviations in the sample and theirs standard deviations for 
distribution of P angles. The applied statistics in details were 
presented in our previous papers
(God{\l}owski et al. 2010 and God{\l}owski 2011a). We compared
results obtained for real sample of our 43 clusters with that
obtained from numerical simulations.

For the sample of all 43 clusters located in superclusters the
distributions of the position angles of members galaxies in the
cluster are anisotropic and the departure from the isotropy is
greater than $3\sigma$ (see Tab. 2 and Tab.3). For the angles which
give the spatial orientation of galaxy planes ($\delta_d$ and $\eta$
angles) the anisotropy is even greater but one should remember the
above problem with aproximation of the "true" ratio of axes of
galaxies as 0.2.

The main point of our study is connected with the trends appearing
in the data. In the analized sample of 43 galaxies we do not observe
the effects connected with cluster richness (Fig.2, Tab.4) which is
significantly different form result obtained by God{\l}owski et al.
2010 for whole sample of $247$ rich Abell clusters. We suppose that
such differences is probably due to environmental efects during
superclusters forming.

\begin{table}[t]
\begin{center}
\scriptsize
\caption{The statistical analysis: value of analyzed statistics for different superclusters richness.}
\label{tab:t5}
\begin{tabular}{ccccc}
\hline
angle&Test&$N=4$&$N=5-7$&$N=8-10$\\
\hline
$P$           &$\chi^2                       $&$43.30 \pm 2.42$&$  34.99 \pm 2.11$&$  36.65 \pm  1.55$\\
              &$\Delta_{1}/\sigma(\Delta_{1})$&$ 1.99 \pm 0.25$&$   1.50 \pm 0.16$&$   1.89 \pm  0.32$\\
              &$\Delta/\sigma(\Delta)        $&$ 2.57 \pm 0.23$&$   2.08 \pm 0.18$&$   2.42 \pm  0.36$\\
\hline
$\delta_D$&$\chi^2_c                         $&$54.52 \pm10.69$&$ 48.68 \pm 4.57$&$   89.07 \pm 18.56$\\
          &$|\Delta_{11}/\sigma(\Delta_{11})|$&$ 3.13 \pm 0.63$&$  2.76 \pm 0.57$&$    6.41 \pm  0.84$\\
          &$\Delta_{c}/\sigma(\Delta_{c})    $&$ 4.76 \pm 0.72$&$  4.40 \pm 0.34$&$    6.97 \pm  1.09$\\

\hline
$\eta$       &$\chi^2                        $&$83.60 \pm12.09$&$  95.57 \pm 9.74$&$  43.01 \pm  5.44$\\
             &$\Delta_{1}/\sigma(\Delta_{1}) $&$ 5.43 \pm 0.86$&$   5.83 \pm 0.71$&$   2.02 \pm  0.29$\\
             &$\Delta/\sigma(\Delta)         $&$ 6.30 \pm 0.88$&$   7.31 \pm 0.66$&$   3.19 \pm  0.38$\\
\hline
\end{tabular}
\end{center}
\end{table}

In the Tab.5 we presented value of analyzed statistics for different
superclusters richness. Generally the anisotropies decreased with
the superclusters richness. Anisotropy for the angles $P$ and
$\delta_d$ seems to increase again for very rich supercluster. One
should note however, that subsample of supercluser richness $8-10$
contain significantly less clusters ($7$) in comparision to the
poorer superclusters and moreover two cluster belonging to this bin
have possible double identifications. Result of linear regresion
between values of analized statistics and supercluster richness
presented on the Fig.3 and in Tab.6 generally confirmed above
conclusion, with exception for $\delta_d$ angle where infuence of
the $8-10$ bin is significant. Because our binned analizis is based
only for 3 bins it is difficult to decide about statistical
significance of this results. For this reason we repeat our analyzis
for not binned sample of 46 clusters (3 of them have possible double
identifications). Analising statistics $T= \frac{a}{\sigma(a)}$
which has Student's distribution with $n-2$ degrees of freedom, we
conclude that only in the case of the $\eta$ angle decreasing the
anisotropy with the supercluster richness is statisticaly
significant on the level of $0.05$.

\begin{figure}
\begin{center}$
\begin{array}{lll}
\includegraphics[angle=270,scale=0.14]{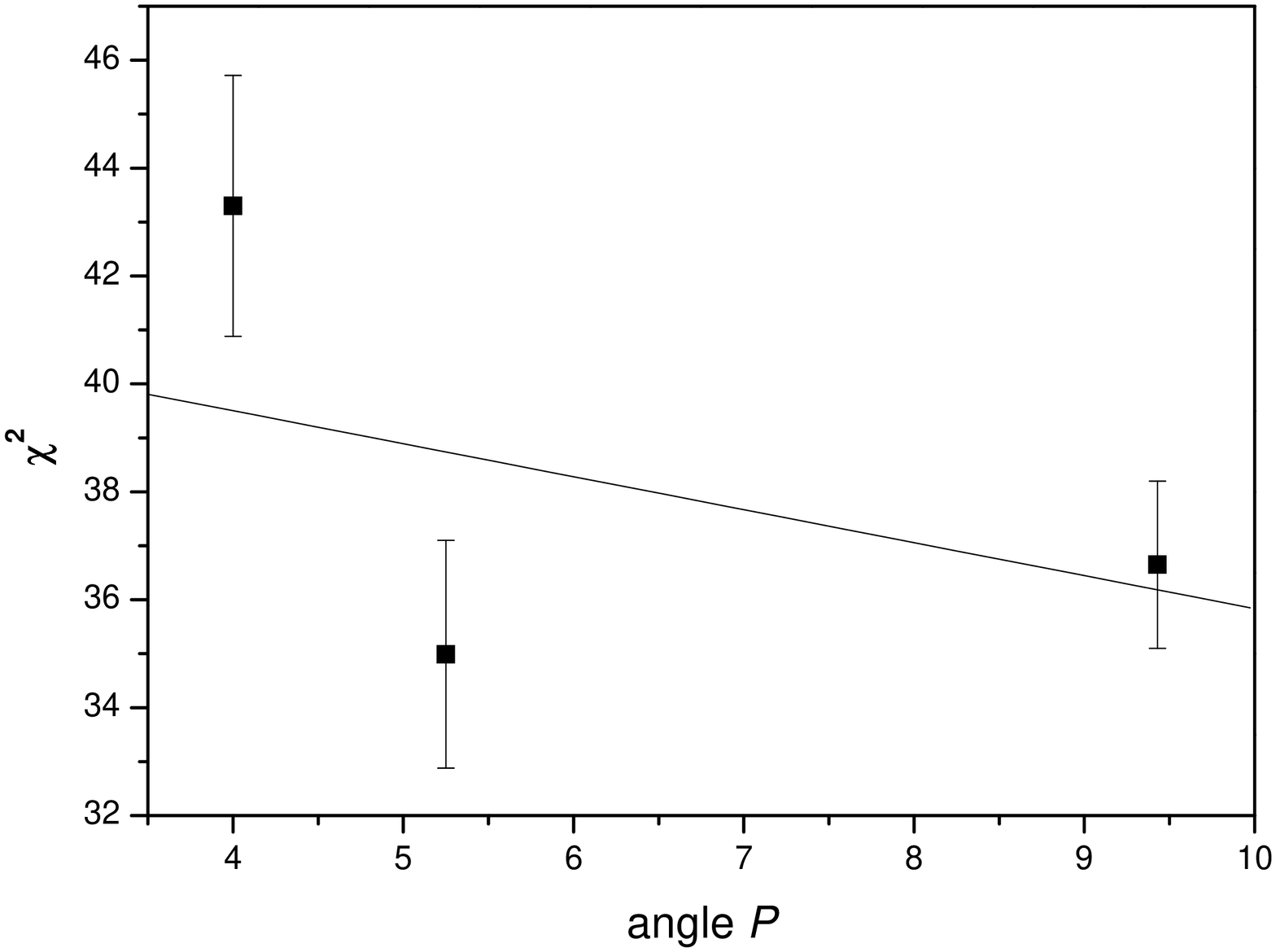} & \includegraphics[angle=270,scale=0.14]{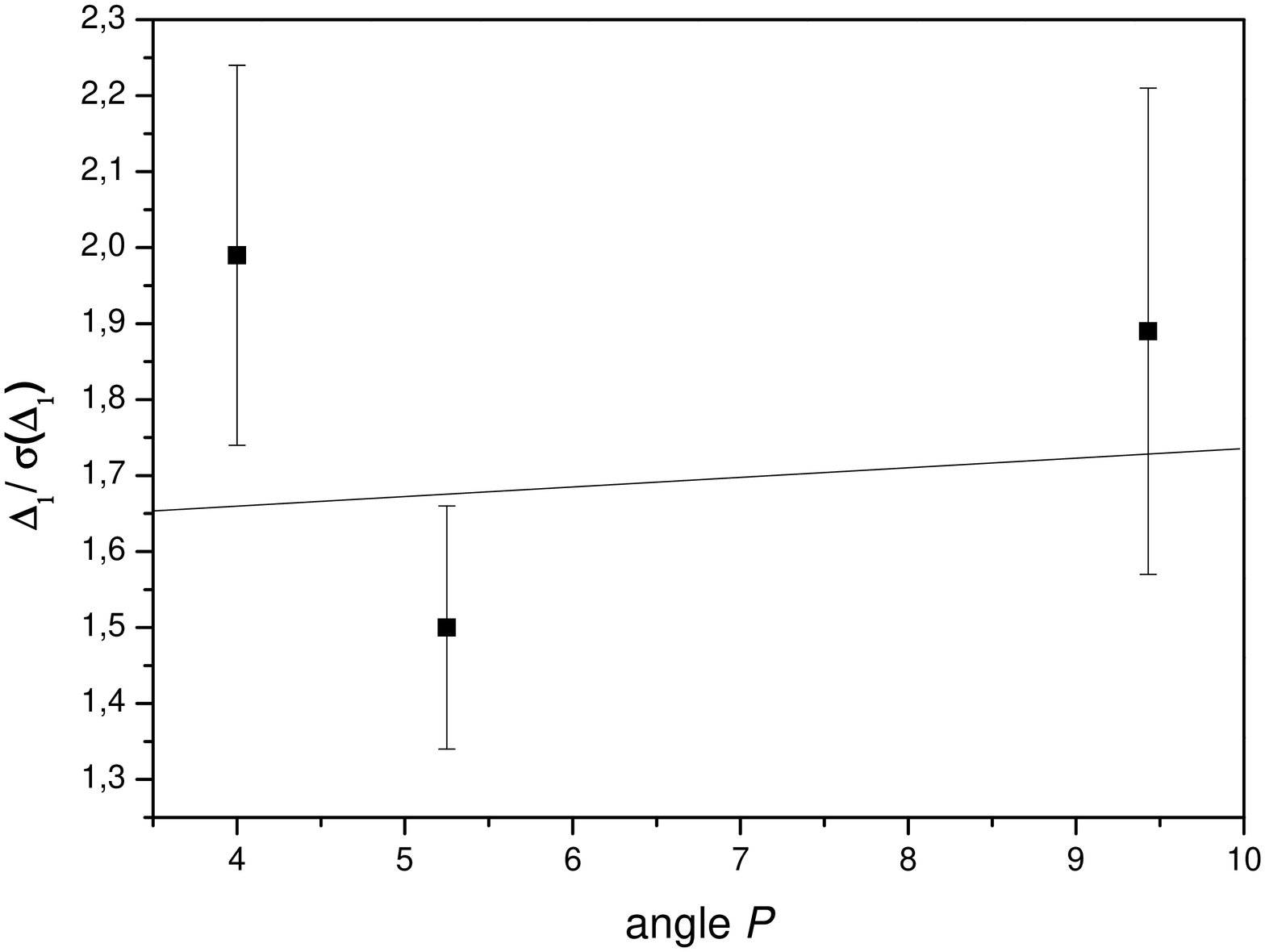} & \includegraphics[angle=270,scale=0.14]{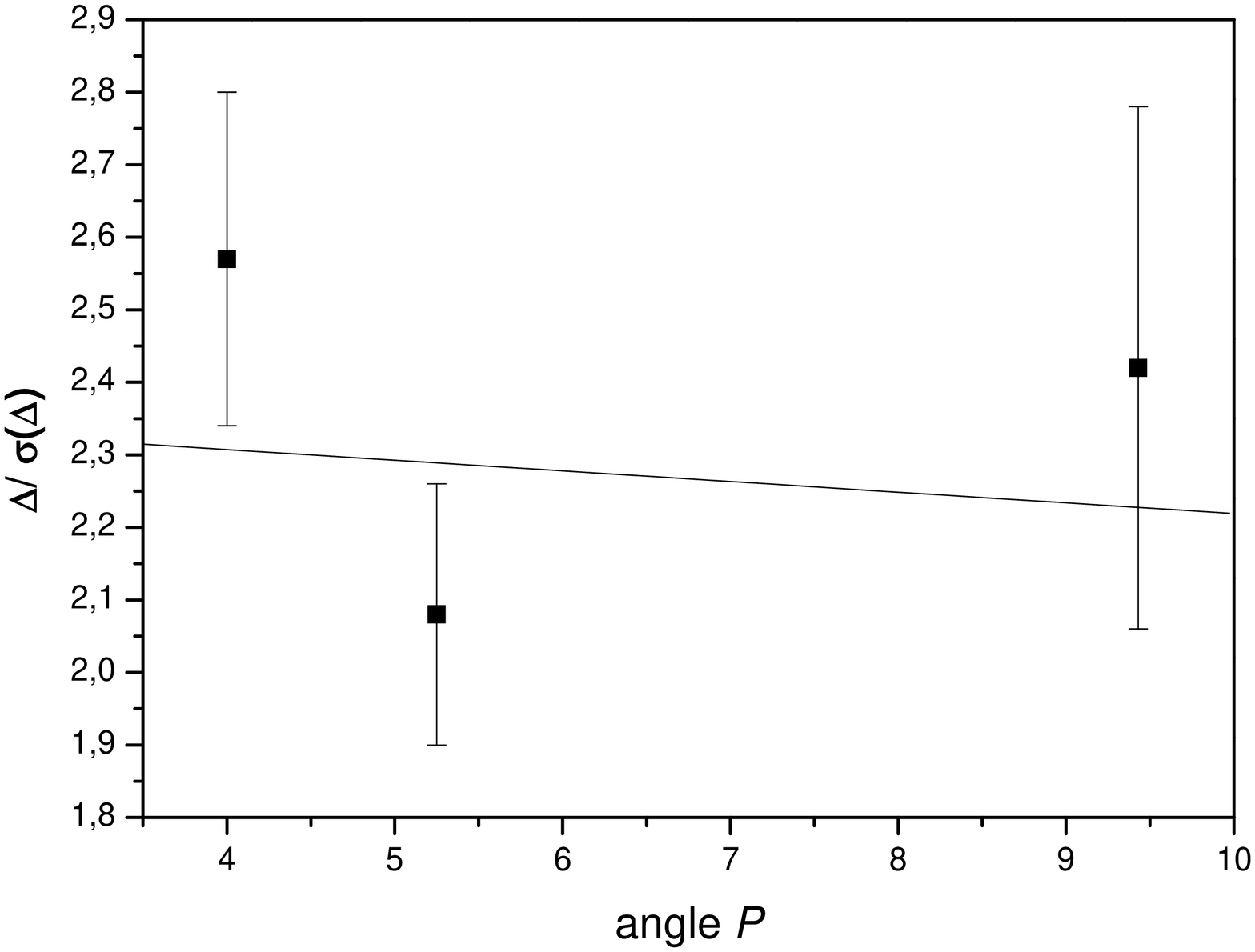}\\
\includegraphics[angle=270,scale=0.14]{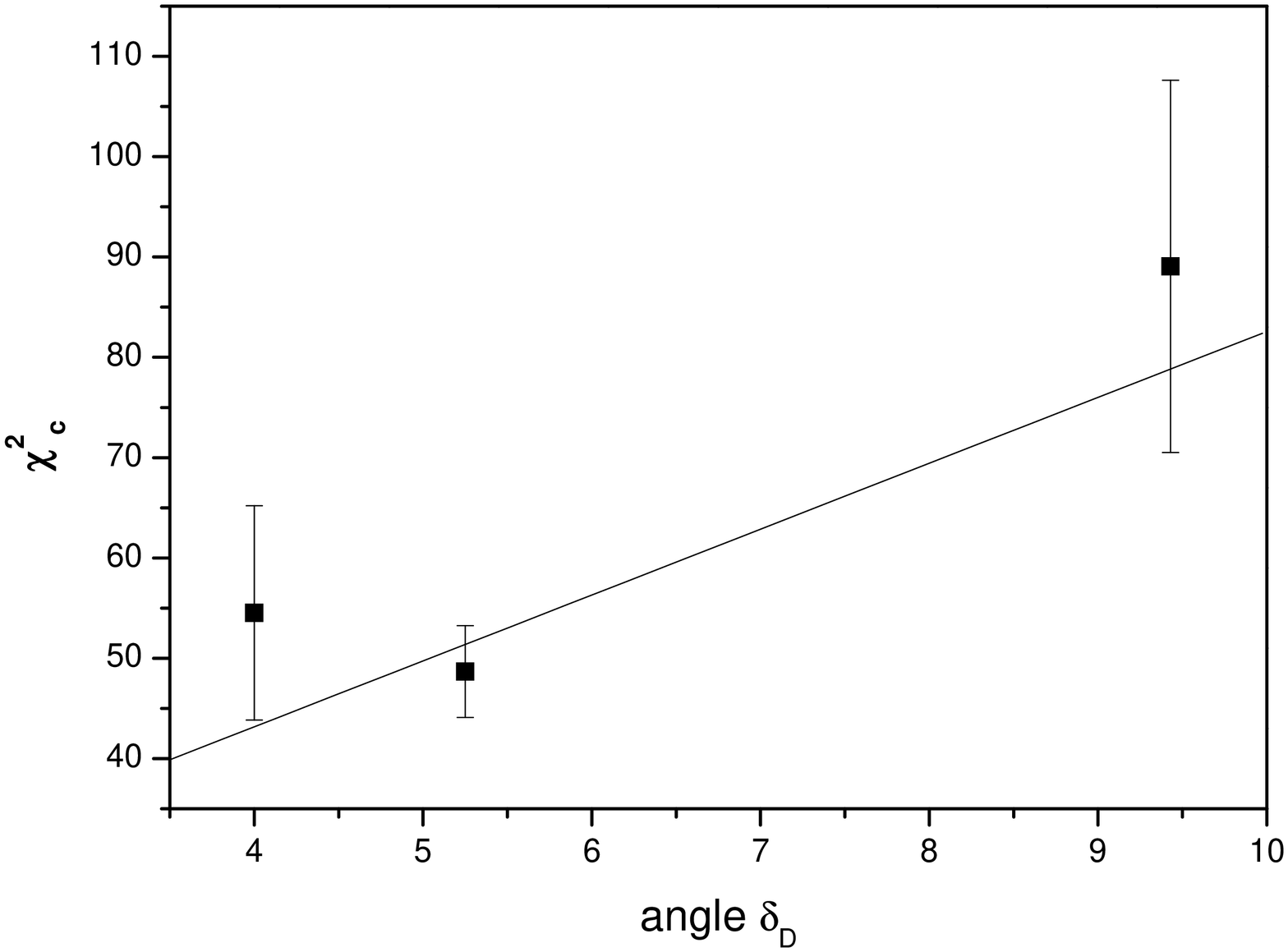} & \includegraphics[angle=270,scale=0.14]{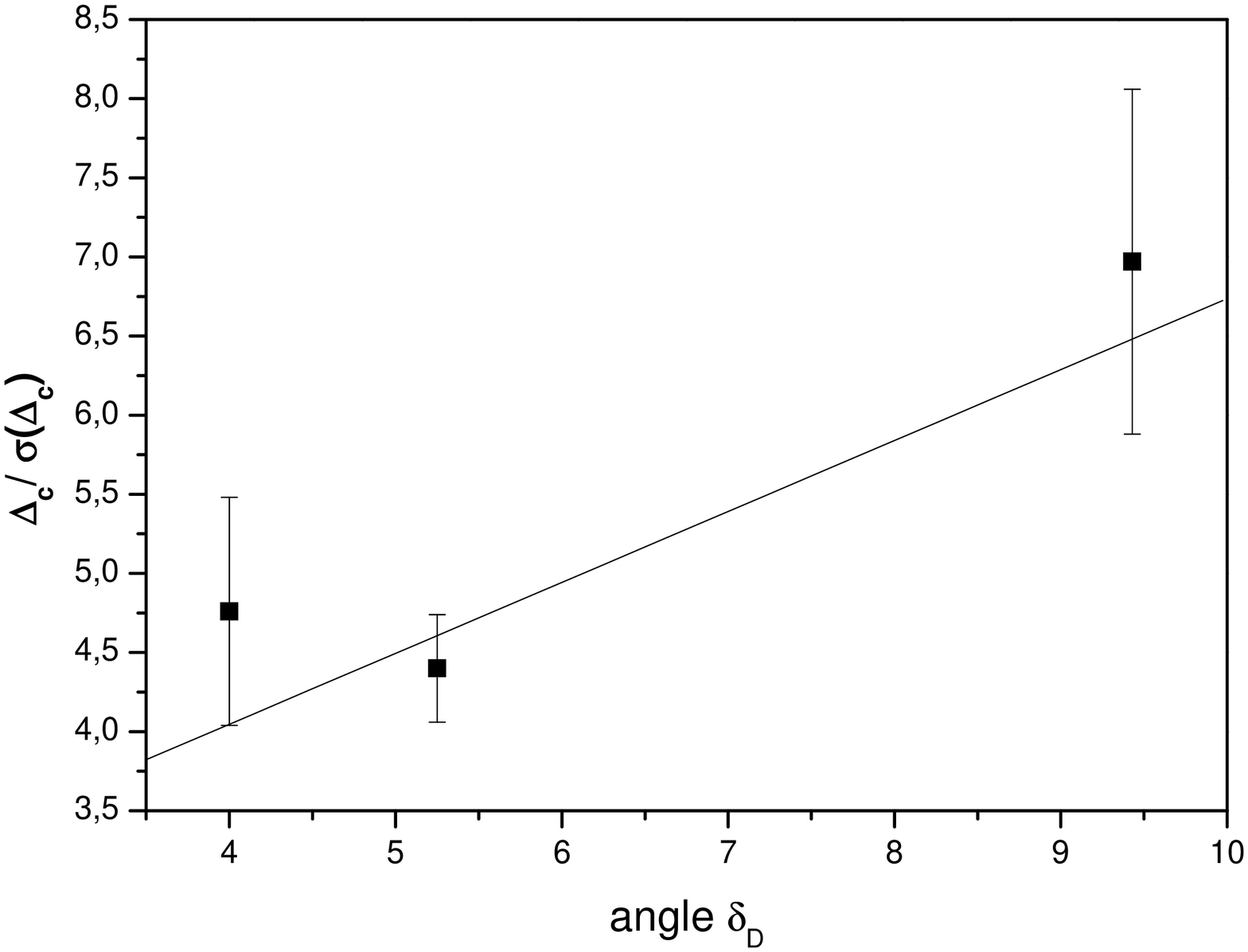} & \includegraphics[angle=270,scale=0.14]{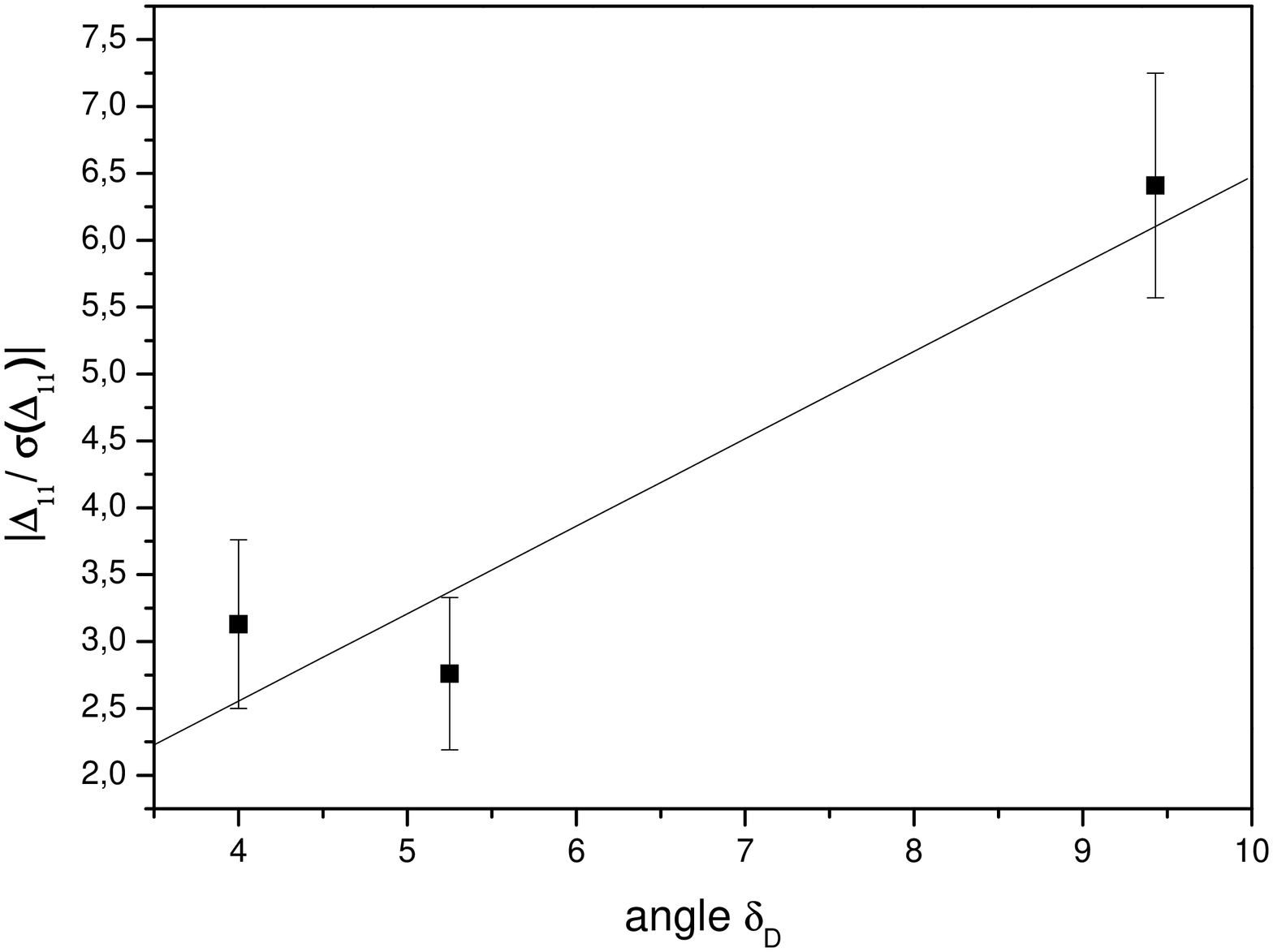}\\
\includegraphics[angle=270,scale=0.14]{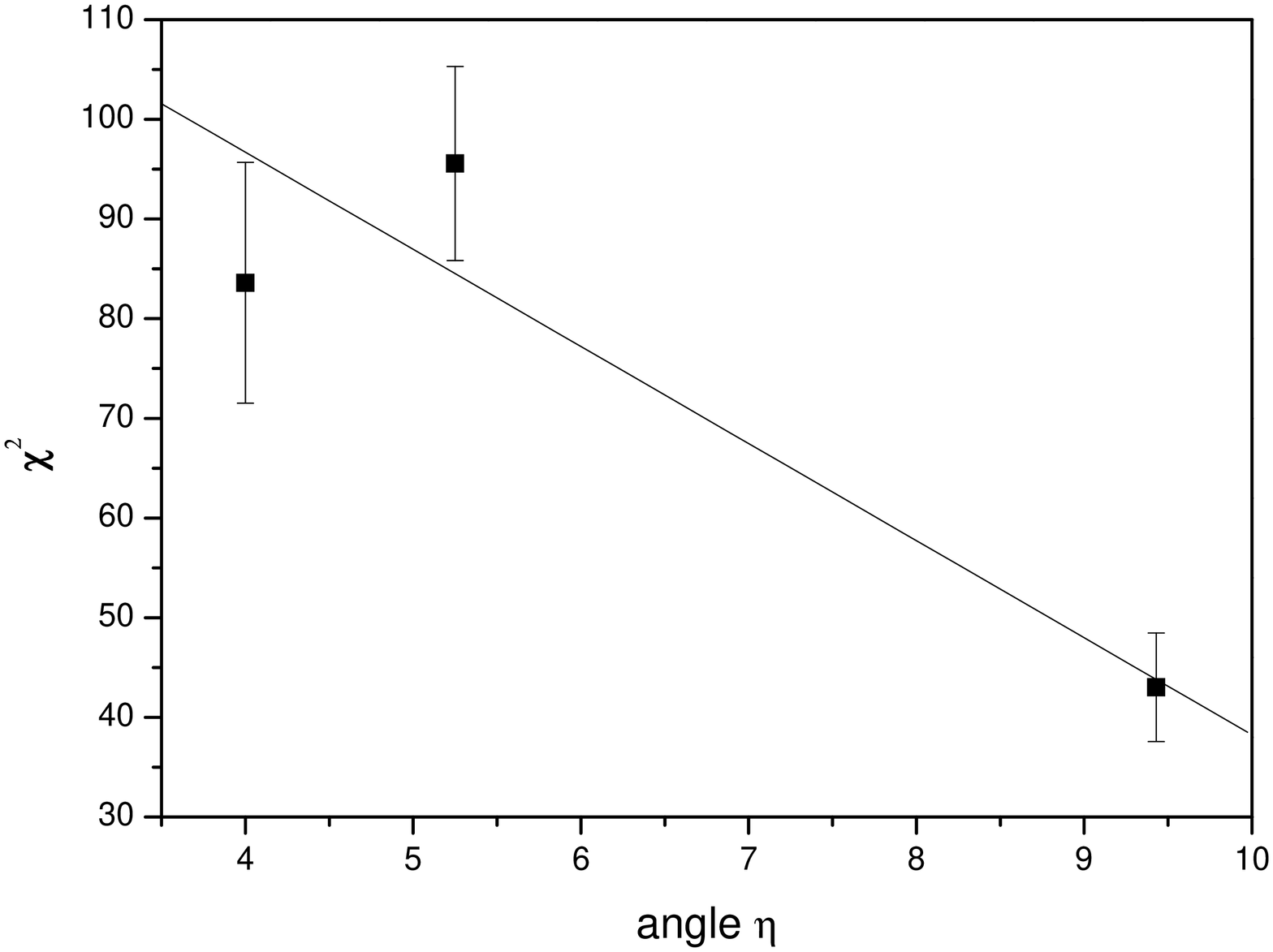} & \includegraphics[angle=270,scale=0.14]{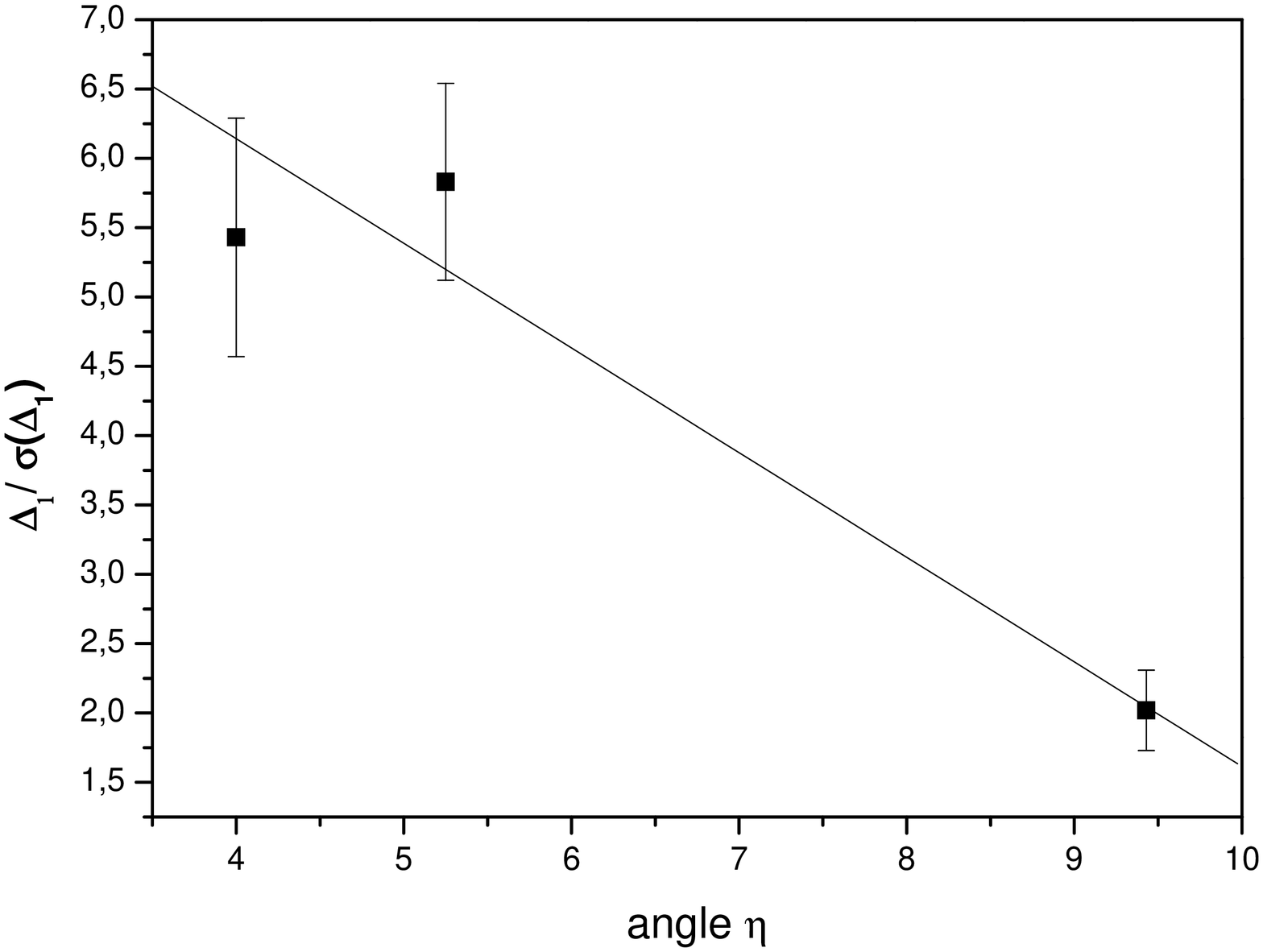} & \includegraphics[angle=270,scale=0.14]{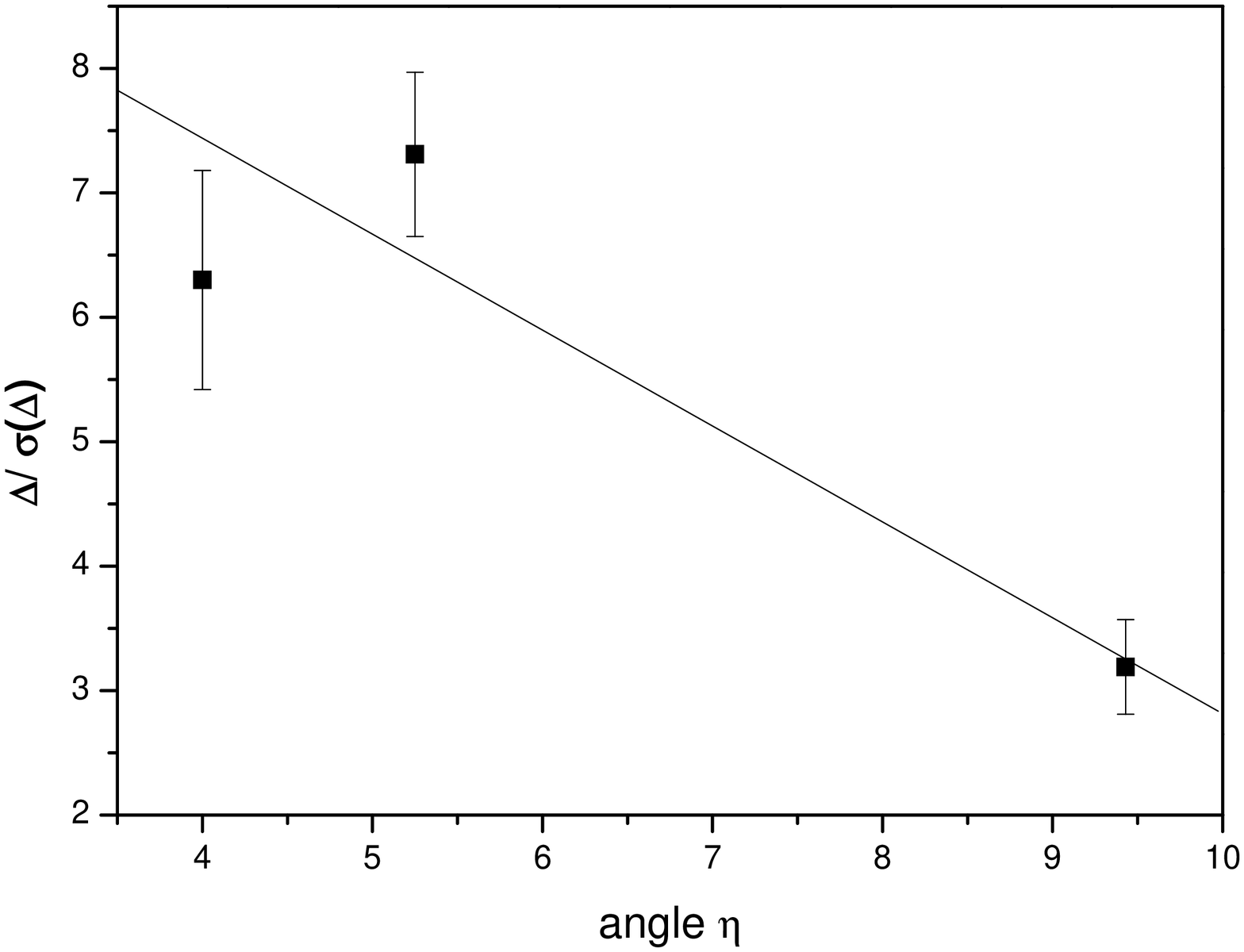}\\
\end{array}$
\end{center}
\caption{The result of the regresion analizis between value of
statistics ($\chi^2$ - left panel, $\Delta_1/\sigma(\Delta_1)$ -
middle panel, $\Delta/\sigma(\Delta)$ - right panel) and the number
of clusters in superscusters for the position angles $p$ (upper
panel),$\delta_D$ angles (midlle panel) and $\eta$ angles (bottom
panel) - equatorial coordinates. \label{fig3}}
\end{figure}

\begin{table}
\begin{center}
\scriptsize \caption{The results of the linear regression analysis -
value of analyzed statistics on the function of supercluster
richness \label{tab:t6}} \resizebox{\textwidth}{!}{
\begin{tabular}{c|cc|cc|cc}
\hline
\multicolumn{1}{c}{}&
\multicolumn{2}{c}{$\chi^2$}&
\multicolumn{2}{c}{$\Delta_1/\sigma(\Delta_1)$}&
\multicolumn{2}{c}{$\Delta/\sigma(\Delta)$}\\
\hline
angle&$a \pm\sigma(a)$&$b\pm\sigma(b)$&$a\pm\sigma(a)$&$b\pm\sigma(b)$&$a\pm\sigma(a)$&$b\pm\sigma(b)$\\
\hline
$P$     &$-.61\pm 0.46$&$41.9\pm3.5 $&$0.014\pm0.074$&$1.61\pm0.43$&$-.016\pm0.080$&$2.39\pm0.45$\\
$\delta$&$6.57\pm 3.94$&$16.9\pm21.2$&$0.656\pm0.191$&$-.08\pm1.14$&$0.445\pm0.240$&$2.27\pm1.31$\\
$\eta$  &$-9.74\pm 1.97$&$135.\pm16. $&$-.756\pm0.131$&$9.14\pm1.13$&$-.770\pm0.139$&$10.5\pm1.1 $\\
\hline
\end{tabular}}
\end{center}
\end{table}

\section{Discussion and conclusions}

We investigated sample of $43$ rich Abell galaxy cluster belonging
to the supercluster and having at least 100 mebers in the considered
area. As expected, the analised superclusters are rather flat
structures. We found that orientation of galaxies in the analised
cluster are not random. God{\l}owski et al. (2011) found that for
the rich Abell cluster alignment increases with the cluster
richness. In contrast, the cluster belonging to the superclusters
does not show such effect. The alignment in poor supercluster is
greater that in the case of rich one. The obtained results, which
show the dependence of galaxy alignment on both the cluster location
inside or outside the supercluster and the supercluster richness
clearly suport the influence of environmental effects to the origin
of galaxy angular momenta. In a very simple and naive picture, if
the alignment of galaxies is primordial, the strongest effect should
be observed in small structures. However, in the present moment we
do not have such data. We considered only really very rich clusters
and looked for trends. It could be accepted that gaining of angular
momenta for galaxies in structure is a rather complicated problem,
in which several mechanism played roles. In some cases the angular
momentum of galaxies is due to local anisotropic collapse of a
protostructures, in other ones due to tidal torque mechanism.
Moreover, cluster merging introduces additional factors influencing
the observed distribution of galaxy angular momenta. This suggests
that the environment played crucial role in origin of galaxy angular
momentum. Our result is preliminary because our investigations were
based on sample of only very rich clusters and will be continued
with analizis of the structures with various richness taken into
account. Fortunately, PF catalogue will allow us to perform such
analysis.


\begin{thebibliography}{}
\bibitem {ACO} Abell, G.O., Corwin, H.G., Jr., Olowin, R.P., 1989, ApJS, 70, 1
\bibitem {Ar00} Aryal, B., Saurer, W., 2000, A\&A, 364, L97
\bibitem {a2} Aryal, B., Neupane, D., Saurer, W., 2008 A\&SS 314, 177
\bibitem {a3} Aryal, B., Paudel, S., Saurer, W., 2008 A\&A, 479, 397
\bibitem {a4} Aryal, B., Saurer, W., 2005 A\&A, 432, 431
\bibitem {Baier03} Baier, F. W., God{\l}owski, W., MacGillivray, H. T. 2003, A\&A, 403,847
\bibitem {Bower05} Bower, R. G., Benson, A.J., Malbon, R., Helly, J., Frenk, C. S., Baugh, C. M., Cole, S., Lacey, C. G. 2006, MNRAS, 370, 645
\bibitem {Ca98} Cabanela, J.E., Aldering, G., 1998 AJ 116, 1094
\bibitem {Catelan96} Catelan, P., Theuns, T. 1996 MNRAS, 282, 436
\bibitem {Djorgovski83} Djorgovski, S. 1983, ApJ, 274, L7
\bibitem {Doroshkevich73} Doroshkevich, A. G. 1973, ApL, 14, 11
\bibitem {Flin88} Flin, P. 1988 MNRAS 235, 857
\bibitem {f01} Flin, P., 2001 MNRAS 325, 49
\bibitem {f4} Flin, P., God{\l}owski, W. 1986, MNRAS, 222, 525
\bibitem {f5} Flin, P., God{\l}owski, W., 1989 Sov. Astron. Lett. 15, 374
 (Pisma w Astro. Zhurnal 15, 867)
\bibitem {f6} Flin, P., God{\l}owski, W., 1990 Sov. Astron. Lett. 65, 209
 (Pisma w Astro. Zhurnal 16, 490)
\bibitem {ga93} Garrido J.L., Battaner, E., Sanchez-Saavedra, M.L., Florido, E., 1993 A\&A 271, 84
\bibitem {g2} God{\l}owski, W. 1993, MNRAS, 265, 874
\bibitem {g3} God{\l}owski, W. 1994, MNRAS, 271, 19
\bibitem {g11} God{\l}owski, W. 2011, IJMPD 20, 1643
\bibitem {g11a} God{\l}owski, W., 2011a, arXiv:1110.2245
\bibitem {g4} God{\l}owski, W., Baier, F.W. MacGillivray, H.T., 1998, A\&A 339, 709
\bibitem {g10} God{\l}owski, W., Flin, P., 2010, ApJ 708, 902
\bibitem {g5} God{\l}owski, W., Ostrowski, M., 1999, MNRAS 303, 50
\bibitem {g10a} God{\l}owski, W., Piwowarska, P., Panko, E., Flin, P., 2010, ApJ 723, 985
\bibitem {gr81} Gregory, S.A., Thompson, L.A., Tifft, W.G., 1981 ApJ 243, 411
\bibitem {h4} Hawley, D. I., Peebles, P. J. E. 1975, AJ, 80, 477
\bibitem {Hu06} Hu F.X., Wu G.X., Song G.X., et al. 2006 A\&SS 302, 42
\bibitem {Jones10} Jones, B., van der Waygaert R., Aragon-Calvo M., 2010 MNRAS, 408, 897
\bibitem {K92} Kashikawa, N., Okamura, S., 1992 PASJ 44, 493
\bibitem {Kindl87} Kindl, A. 1987, AJ 93, 1024
\bibitem {Lee02} Lee, J., Pen, U. 2002, ApJ, 567, L111
\bibitem {Li98} Li, Li-Xin., 1998, Gen. Rel. Grav., 30, 497
\bibitem {MG82} MacGillivray, H.T., Dodd, R.J., McNally, B.V., Corwin, Jr. H.G., 1982, MNRAS, 198, 605
\bibitem {Navarro04} Navarro, J. F., Abadi, M. G., Steinmetz M. 2004, ApJ, 613, L41
\bibitem {Panko06} Panko, E., Flin, P., 2006, Journ. Astro. Data 12,1
\bibitem {Peebles69} Peebles, P.J.E. 1969, ApJ, 155, 393
\bibitem {Shandarin74} Shandarin, S.F. 1974, Sov. Astr. 18, 392
\bibitem {Silk83} Silk, J., Efstathiou, G. A. 1983, The Formation of Galaxies, Fundamentals of Cosm. Phys. 9, 1
\bibitem {Sunyaew72} Sunyaev, A. R., Zeldovich, Ya. B., 1972 A\&A, 20, 189
\bibitem {t06} Trujillo, I., Carretro, C., Patiri, S.G., 2006, ApJ 640, L111
\bibitem {t88} Tully, R. B., Nearby Galaxy Catalog, Cambridge 1988
\bibitem {MRSS03} Ungruhe, R., Seitter, W., Durbeck, H., 2003, Journ. Astr. Data, 9.1
\bibitem {v11} Varela, J. Rios,J.B., Trujillo, I., 2011, astro-ph 1109.2056
\bibitem {Wesson82} Wesson, P. S. 1982, Vistas Astron., 26, 225
\bibitem {Wu97} Wu G.X., Hu F.X., Su H.J., Liu Y.Z., 1997 A\&A, 323, 317
\bibitem {Zeldovich70} Zeldovich, B. Ya. 1970, A\&A, 5, 84
\end{thebibliography}
\end{document}